\documentclass[prb,aps,amsmath,amssymb,showpacs,twocolumn,notitlepage]{revtex4-1}
\usepackage{amssymb,graphicx,float,dcolumn,bm,subfigure,color}
\usepackage[titletoc,title]{appendix}
\usepackage[dotinlabels]{titletoc}
\usepackage{subfigure}
\usepackage{braket}
\usepackage{nicefrac}
\usepackage[dvipsnames]{xcolor}

\newcommand{\be}{\begin{equation}}
\newcommand{\ee}{\end{equation}}

\newcommand{\ba}{\begin{eqnarray}}
\newcommand{\ea}{\end{eqnarray}}

\renewcommand{\vec}[1]{{\textbf{\textit{#1}}}}

\begin{document}
\title{It's anyon's game: the race to quantum computation}
\author{Jainendra K. Jain}
\affiliation{Physics Department, 104 Davey Lab, Pennsylvania State University, University Park, Pennsylvania 16802, USA}
\begin{abstract}
In 1924, Satyendra Nath Bose dispatched a manuscript introducing the concept now known as Bose statistics to Albert Einstein. Bose could hardly have imagined that the exotic statistics of certain emergent particles of quantum matter would one day suggest a route to fault-tolerant quantum computation. This non-technical Commentary on ``anyons," namely particles whose statistics is intermediate between Bose and Fermi, aims to convey the underlying concept as well as its experimental manifestations to the uninitiated.
\end{abstract}

\maketitle

\pagebreak

{\em Fractional statistics:} 
The quantum mechanical state of a collection of particles is described by a complex valued wave function $\Psi(\{\vec{r}_j\})$, with $|\Psi(\{\vec{r}_j\})|^2$ giving the probability that the particles are located at positions $(\vec{r}_1, \vec{r}_2, \cdots \vec{r}_N )$.  An exchange of two identical particles does not produce a new state, and in particular, leaves the probability $|\Psi(\{\vec{r}_j\})|^2$ unchanged. This imposes certain symmetry constraints on the allowed wave functions. Two possibilities are $\Psi\rightarrow + \Psi$ or $\Psi\rightarrow - \Psi$ under exchange of two particles. Particles obeying the former relation are called bosons, and the latter fermions.  We speak of Bose and Fermi statistics because the property of particles under exchange affects the counting of distinct microscopic configurations 
and thus their statistical mechanics and thermodynamics.

Examples of fermions are electrons, quarks, protons, neutrons and $^3$He atoms; examples of bosons are the Higgs bosons, $^4$He atoms and photons. Even though particle statistics may appear an abstract concept unconnected to our daily lives, it is central to our existence. The Fermi statistics of electrons is responsible for the structure of atoms and molecules, and thus permeates atomic physics, chemistry and biology; it governs, at the most fundamental level, the properties of all matter around and within us. The Bose statistics is responsible for dramatic phenomena at low temperatures, such as Bose-Einstein condensation, superconductivity and $^4$He superfluidity.

\begin{figure}[t]
\includegraphics[width=3.3in, scale=1]{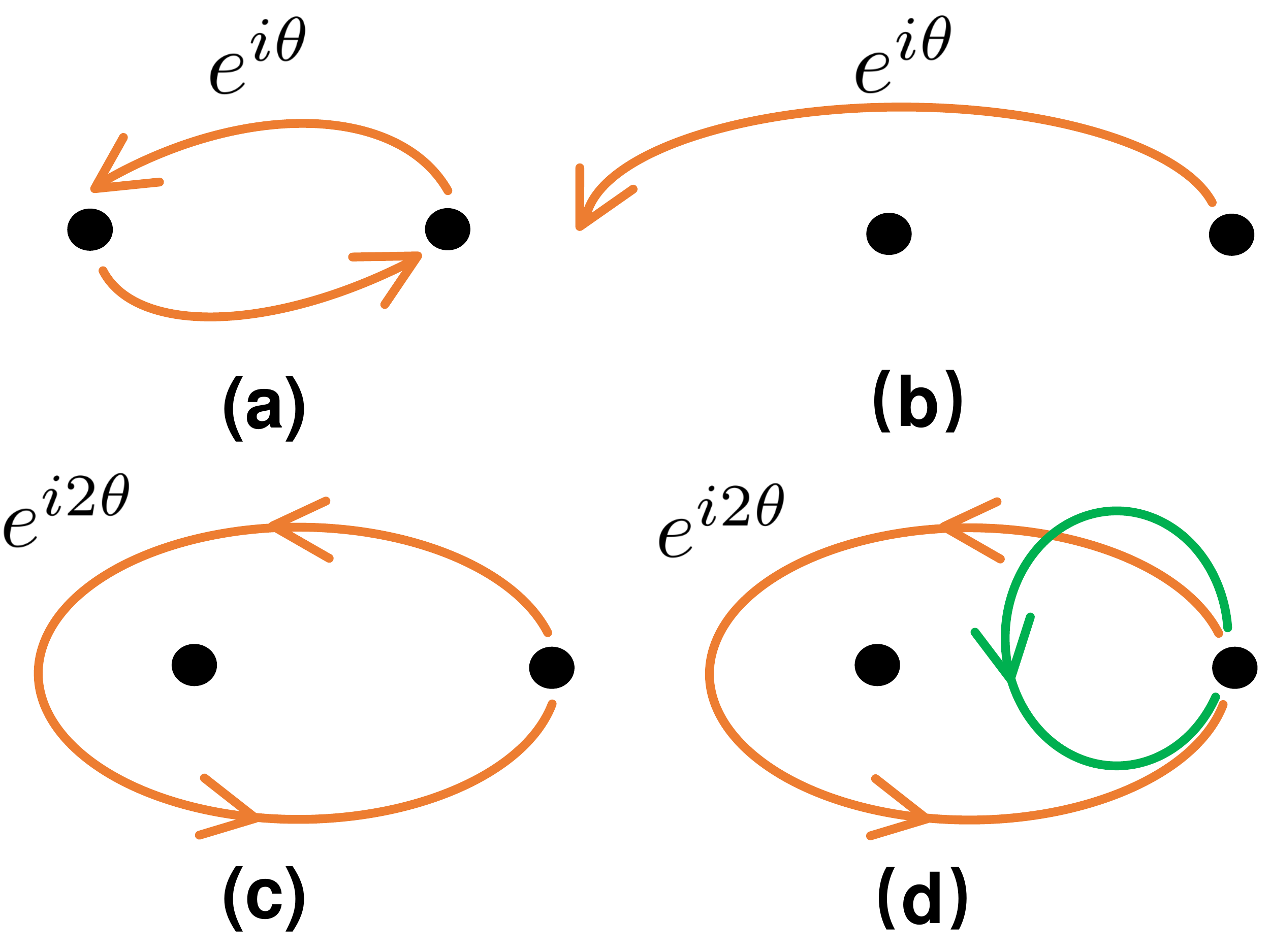}
\caption{(a) Exchange of two particles produces a phase factor $e^{i\theta}$. (b) Particle exchange is equivalent to half a loop of one particle around another. (c) A full loop equals two exchanges. (d) In two dimensions the orange and green paths are topologically distinct, but in higher dimensions they are not.}\label{Fig1}
\end{figure}

It was recognized four decades ago~\cite{Leinaas77,Wilczek82} that, in two space dimensions, quantum mechanics also admits particles that behave under exchange as $\Psi\rightarrow e^{i\theta} \Psi$ with arbitrary $\theta$. To see the essential idea, let us imagine exchanging two particles along a counterclockwise path shown in Fig.~\ref{Fig1} (a), which produces a phase factor $e^{i\theta}$. (A clockwise exchange will produce $e^{-i\theta}$.) From the reference frame of one of the particles, this is a half loop [Fig.~\ref{Fig1} (b)]. A full loop [Fig.~\ref{Fig1} (c)], called a winding or braiding, represents two exchanges. The phase $\theta$ is called the (braiding) statistical phase. For bosons (fermions) we have $\theta=m\pi$ where $m$ is an even (odd) integer. Particles for which $\theta$ is not an integer multiple of $\pi$ are referred to as anyons, which stands for particles with any statistics. It is stressed that $\theta$ is a topological quantity, i.e. it does not depend on the size and shape of the exchange path or the closed loop. Anyons can be defined only in two space dimensions, because in higher dimensions the notion of a counterclockwise / clockwise exchange or of a particle going around another is not well defined. In three dimensions, by lifting the paths off the page, one can continuously deform the counterclockwise exchange into a clockwise one in Fig.~\ref{Fig1} (a), or the orange loop into the green loop in Fig.~\ref{Fig1} (d); this implies $e^{i\theta}=e^{-i\theta}$ and hence $\theta=m\pi$. 

Of course, just because anyons can be defined does not mean that they exist. Our world is three dimensional, and there certainly are no anyons in the list of particles in a particle physics text book. Fortunately, this list is incomplete. Interacting quantum matter generates its own emergent particles, which can be rather exotic, and no principle of physics precludes the possibility that  some of them might obey fractional braiding statistics. A quantum state supporting such particles would have to be a rather exotic state in two dimensions. Soon after the concept of fractional statistics was proposed, nature produced a promising candidate for its realization~\cite{Halperin84,Arovas84}, namely the fractional quantum Hall effect. 

{\em Fractional quantum Hall effect:} Typically, electric current flows in the direction of an applied voltage. In the presence of a magnetic field $B$, however, it flows at an angle; that is, there is a voltage in the direction of the current flow as well as across it. The Hall resistance $R_H$ is the ratio of transverse voltage to current. The laws of classical electrodynamics tell us that $R_H$ is proportional to $B$, as seen routinely by experiments. 

Dramatic quantum phenomena are revealed, however, when electrons are confined to two dimensions, cooled to near zero Kelvin, and subjected to a strong magnetic field $B$. The primary observation is that as $B$ is varied, the Hall resistance exhibits a series of plateaus precisely quantized at $R_H=h/(\nu e^2)$, where $h$ is the Planck's constant, $e$ is electron's charge, and $\nu$ is either an integer or a fraction. The appearance of such simple and universal values that are utterly oblivious to all of the complexities of the sample is an amazing result that has fascinated physicists for decades. 

The observation of integer and fractional values of $\nu$ are referred to as the integer quantum Hall effect~\cite{Klitzing80} (IQHE) and fractional quantum Hall effect~\cite{Tsui82} (FQHE), respectively. The IQHE can be understood using standard methods, because it occurs in a theoretical model of noninteracting electrons, and one can make convincing arguments that the interaction does not destroy it. In contrast, the FQHE results fundamentally due to the interaction between electrons, which produces certain complex liquids of strongly correlated electrons. Close to one hundred such FQH liquids, distinguished by $\nu$, have been observed so far in a variety of two-dimensional materials, such as semiconductor quantum wells and graphene.

\begin{figure}[t]
\includegraphics[width=3.3in, scale=1]{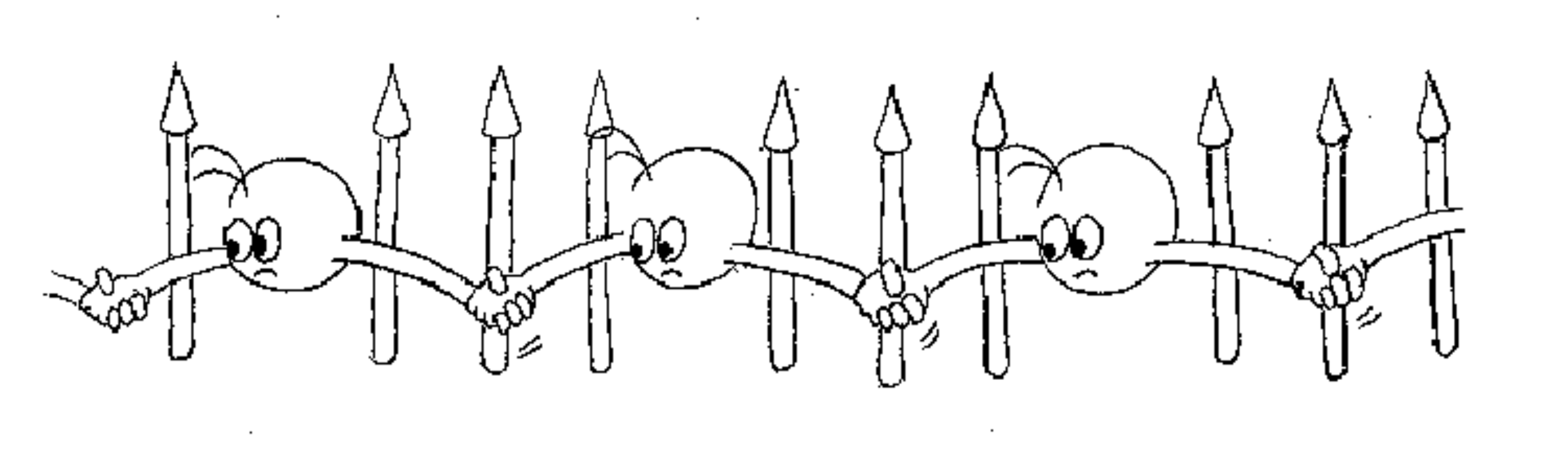}
\includegraphics[width=3.3in, scale=1]{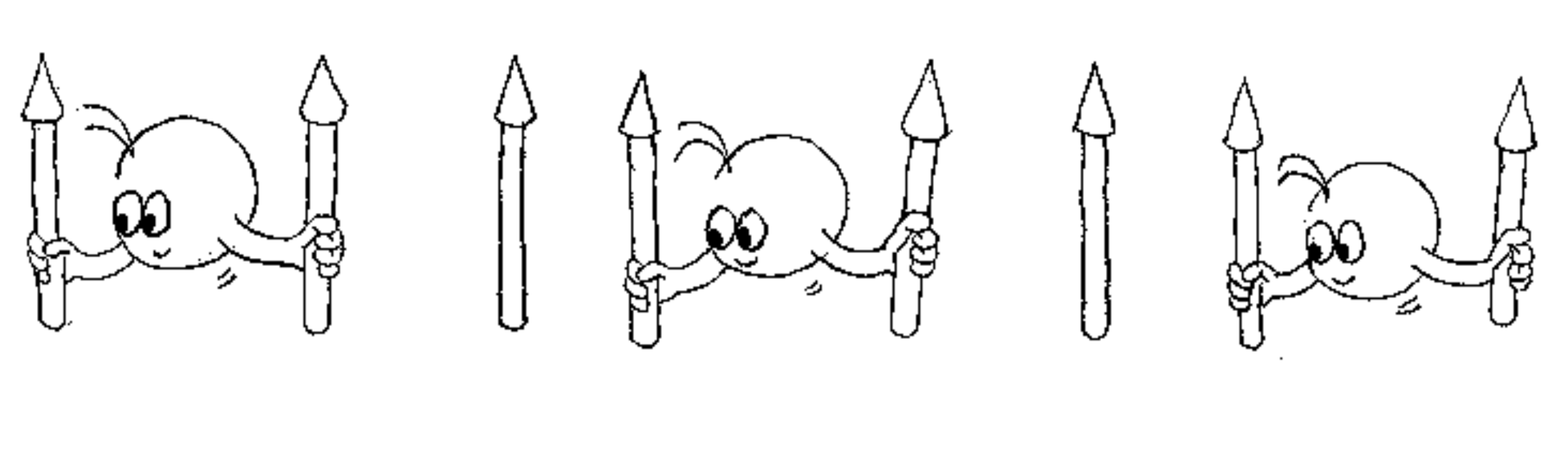}
\caption{A light-hearted depiction of how interacting electrons in a magnetic field (upper row) capture magnetic flux quanta to turn into noninteracting composite fermions (lower row). Illustration by Kwon Park.
\label{Park}}
\end{figure}

{\em Composite fermions:}  
There are several ways of viewing the FQHE. The simplest understanding of its origin is achieved in terms of an emergent particle called the composite fermion~\cite{Jain07}. The composite fermion is an unusual particle: it is often visualized as the bound state of an electron and an even integer number ($2p$) of magnetic flux quanta, where a flux quantum is defined as $\phi_0=h/e$. (This picture of composite fermion is not to be taken literally, but is sufficient for many purposes.) Tremendous simplification occurs because while electrons are strongly interacting, composite fermions are, to a good approximation, noninteracting [Fig.~\ref{Park}]. In other words, the primary role of the interaction between electrons in the FQH regime is simply to create composite fermions. (How many flux quanta composite fermions capture depends on $B$.) Many predictions of  composite fermions have been confirmed. In particular, the IQHE of composite fermions appears as the FQHE of electrons at fractions $\nu=n/(2pn\pm 1)$, $n$ and $p$ being integers, which explains almost all of the observed fractions. Composite fermions have been directly observed, and provide a natural framework for understanding various properties of the FQH liquids.

\begin{figure}[t]
\includegraphics[width=3.3in, scale=1]{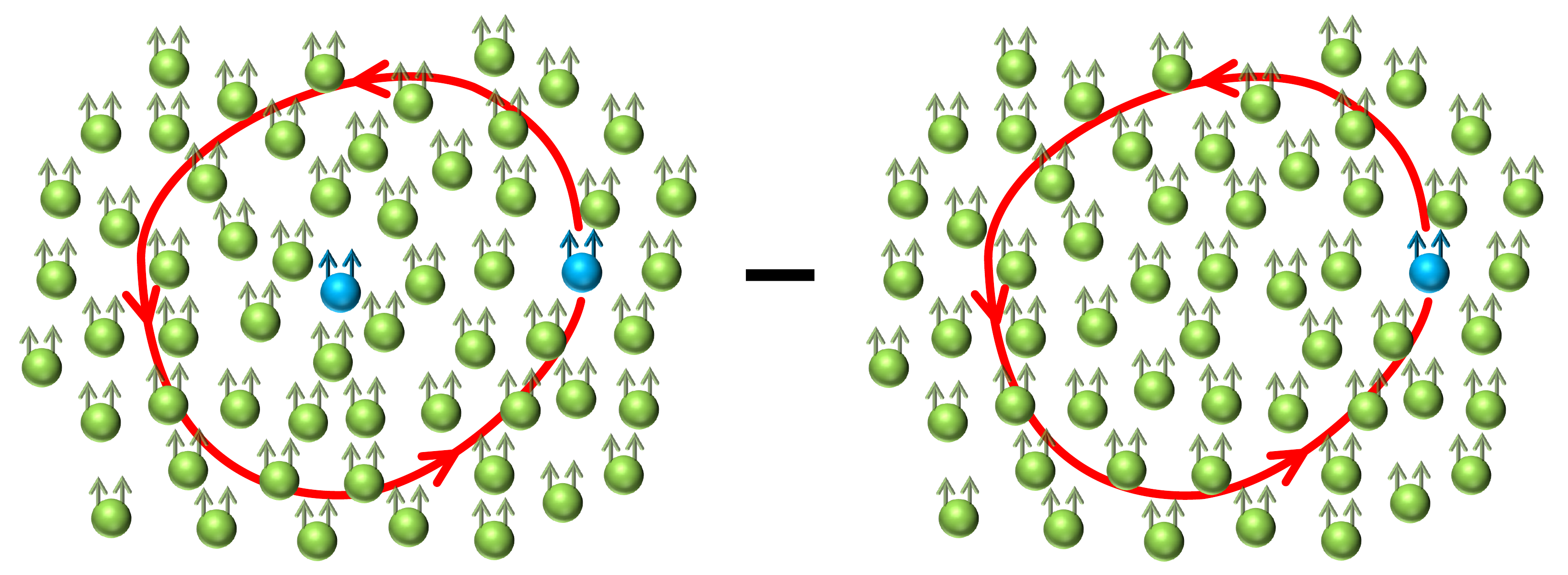}
\caption{Extracting the statistical phase for quasiparticles, i.e. excited composite fermions, shown in blue. The composite fermions belonging to the ground state are shown in green. The statistical phase for the quasiparticles is given by the phase associated with loop on the left, which encloses another quasiparticle, minus the phase associated with the loop on the right, which does not. 
\label{Fig4}}
\end{figure}

\begin{figure}[b]
\includegraphics[width=3.3in, scale=1]{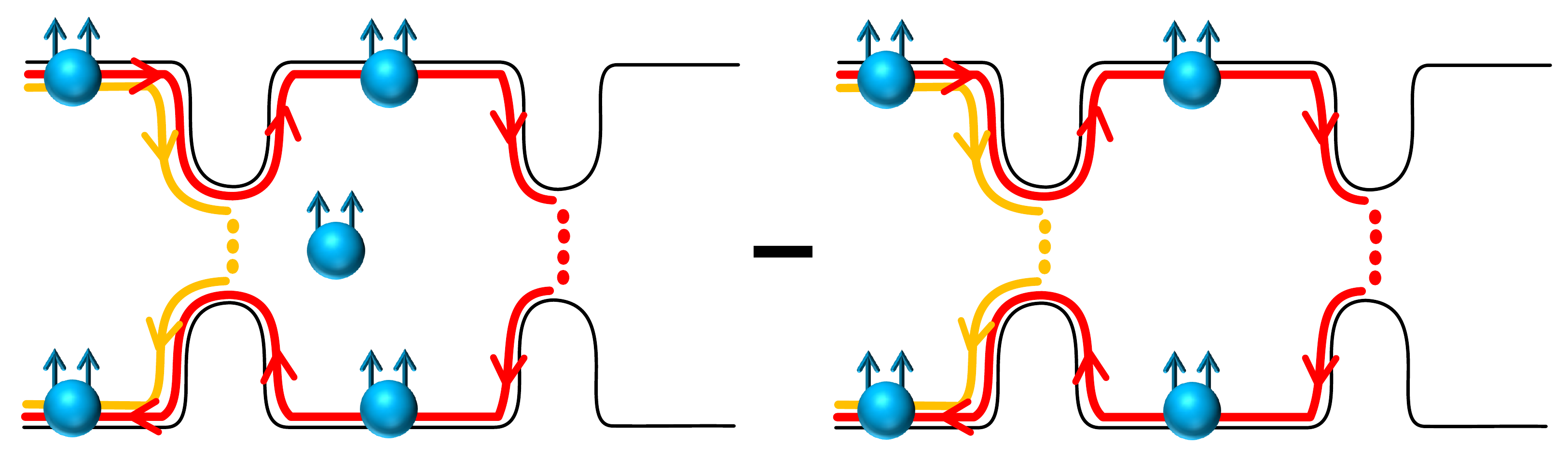}
\caption{Experimental realization of the scheme in Fig.~\ref{Fig4}, using interference of quasiparticles moving along the sample edges. Quasiparticles are injected at the upper left edge, and measured at the lower left edge. The two relevant paths are shown in red and orange, artificially displaced in the region where they coincide. The vertical dotted lines indicate quantum mechanical tunneling across a barrier. Composite fermions belonging to the ground state are not shown to avoid clutter.
}\label{NewFig4}
\end{figure}

What about fractional statistics? The excitations of the FQH liquids are nothing but excited composite fermions, referred to as quasiparticles. It turns out that these quasiparticles obey fractional braiding statistics. The gist of the idea is as follows. Consider a closed loop of a quasiparticle, shown in blue in Fig.~\ref{Fig4}, winding around another quasiparticle. The phase associated with the loop can be calculated, but it is complicated: in addition to the statistical phase $2\theta$, it also involves a contribution from
 the composite fermions in the ground state (shown in green) as well as a phase (called the Aharonov-Bohm phase) that a charged particle acquires when it moves in the presence of a magnetic field. Fortunately, these other path-dependent contributions can be eliminated by subtracting the phase for an identical loop that does not enclose any quasiparticle (see Fig.~\ref{Fig4}).  A microscopic calculation~\cite{Jain07} gives $2\theta=4\pi p/(2pn\pm 1)$ modulo $2\pi$ for the quasiparticles of the $\nu=n/(2pn\pm 1)$ FQH state, consistent with earlier results~\cite{Halperin84,Arovas84}.

{\em Experimental evidence:} To measure the statistical phase experimentally, one first needs to produce well defined paths. Here another property of quantum Hall effect comes in handy: the current flows along the edges of the sample, and moreover, on a given edge it flows only in one direction. This suggests the geometry shown in Fig.~\ref{NewFig4}. A composite fermion  injected into the sample at the upper left edge has several choices. The most probable outcome is that it continues to ride the upper edge and escapes to the right. However, it can also tunnel into the lower edge where the two edges come close. The probability of its appearing at the lower left edge depends, according to the laws of quantum mechanics, on the phase difference between the red and the orange paths, which is precisely the phase associated with the loop encircling the inner island. The change in this phase when a quasiparticle is added to the interior will give the statistical phase $2\theta$, as discussed above. 

Impressive progress has been made by several experimental groups toward detecting fractional braiding statistics using this geometry~\cite{Willett09,An11,Mcclure12,Nakamura19}. One complicating factor is that if the paths are too long, the composite fermion forgets its phase along the way, whereas if they are too short, another phenomenon called Coulomb blockade comes into play. Overcoming these obstacles with a clever sample design, a recent experiment~\cite{Nakamura20} has reported the most convincing observation so far of discrete phase slips of $-2\pi/3$ at $\nu=1/3$, precisely as expected from fractional braiding statistics for the quasiparticles of the $\nu=1/3$ FQH state. 

{\em Quantum computation with anyons?} As physicists work further to better understand, confirm, and probe fractional statistics for the excitations of various FQH states, and search for other phenomena arising from them, an outsider may wonder: How will fractional statistics affect my life? While it would be unwise to make predictions, because new ideas often surprise us by opening unanticipated directions, a dream application would be topological quantum computation~\cite{Nayak08}. This would actually require more complex anyons than the ones discussed above, called ``non-Abelian anyons." For such anyons the outcome of a sequence of windings depends on the order in which they are performed. A special case of such anyons, called Majorana particles, are theoretically believed to occur in the $\nu=5/2$ FQH state wherein composite fermions pair up to form a superconductor~\cite{Read00}. 
This state has also been investigated in interference experiments~\cite{Willett09,An11,Willett13}. Intense search for Majorana particles is also ongoing in contexts outside of the FQHE. How can they help with quantum computation? A fundamental impediment to quantum computation is decoherence due to interaction with the environment. Here, one can imagine a topological qubit (quantum bit) in which information is stored non-locally over two distant Majoranas, and thus immune to decoherence by local fluctuations in the environment. The realization of a quantum computer based on non-Abelian anyons of the FQHE would be a fitting legacy of fractional statistics. 

I thank Yayun Hu for discussions. Financial support from the US Department of Energy, Office of Basic Energy Sciences, under Grant No. DE-SC0005042 is gratefully acknowledged.

\end{document}